\documentclass[aps,prb,superscriptaddress,floatfix,twocolumn,amsmath,amssymb,showpacs]{revtex4-1}
\usepackage{graphicx}
\usepackage{bm}
\usepackage{color}
\usepackage{epstopdf}
\begin{document}
\title{
Tuning the polarized quantum phonon transmission in graphene nanoribbons 
}

\author{P. Scuracchio}
\affiliation{
Facultad de Ciencias Exactas Ingenier{\'\i}a y Agrimensura, Universidad Nacional de Rosario and 
Instituto de F\'{\i}sica Rosario, Bv. 27 de Febrero 210 bis, 2000 Rosario,
Argentina}

\author{A. Dobry}
\affiliation{
Facultad de Ciencias Exactas Ingenier{\'\i}a y Agrimensura, Universidad Nacional de Rosario and 
Instituto de F\'{\i}sica Rosario, Bv. 27 de Febrero 210 bis, 2000 Rosario,
Argentina}

\author{F. M. Peeters}
\affiliation{Universiteit Antwerpen, Department of Physics, Groenenborgerlaan 171, 2020 Antwerpen, Belgium 
}

\author{S. Costamagna}
\affiliation{
Facultad de Ciencias Exactas Ingenier{\'\i}a y Agrimensura, Universidad Nacional de Rosario and Instituto de F\'{\i}sica Rosario, Bv. 27 de Febrero 210 bis, 2000 Rosario,
Argentina}
\affiliation{Universiteit Antwerpen, Department of Physics, Groenenborgerlaan 171, 2020 Antwerpen, Belgium 
}
\email{costamagna@ifir-conicet.gov.ar}
\date{\today}


\begin{abstract}
We propose systems that allow a tuning of the phonon transmission function T($\omega$) 
in graphene nanoribbons by using C$^{13}$ isotope barriers, antidot structures, 
and distinct boundary conditions. 
Phonon modes are obtained by an interatomic fifth-nearest neighbor force-constant model (5NNFCM) 
and T($\omega$) is calculated using the non-equilibrium Green's function formalism. 
%
We show that by imposing partial fixed boundary conditions it is possible to restrict contributions of
the in-plane phonon modes to T($\omega$) at low energy.
On the contrary, the transmission functions of out-of-plane phonon modes can be diminished by 
proper antidot or isotope arrangements. In particular, we show that a periodic array of them 
leads to sharp dips in the transmission function at certain frequencies $\omega_{\nu}$ which can be pre-defined 
as desired by controlling their relative distance and size.
With this, we demonstrated that by adequate engineering it is possible to govern 
the magnitude of the ballistic transmission functions T$(\omega)$ in graphene nanoribbons.
We discuss the implications of these results in the design of controlled thermal transport at the nanoscale
as well as in the enhancement of thermo-electric features of graphene-based materials. 
\end{abstract}



\maketitle

{\it Introduction}.
\label{intro}
Immediately after the first measurements of graphene's extraordinary large thermal 
conductivity\cite{Balandin1, Seol, Ghosh}, its vibrational properties have  
become an object of intense research\cite{Balandin2, Fengi, Balandin3, nika2}. 
The low density of states at the Fermi energy makes contributions 
of free electrons (Dirac fermions) to be negligible~\cite{Novo}
and the intrinsic mechanisms behind the outstanding thermal transport properties 
are then almost completely attributed to the phonon characteristics.
These investigations, together with the 
improvements in fabrication techniques\cite{fab1, fab2},
have put forth the high potential of using graphene in  
obtaining rapid thermal dissipation, a topic highly important 
for present and future nano-electronic devices\cite{Pop}.
Measurements of thermal transport in the ballistic limit, where the phonon mean free-path 
is larger than the dimensions of the sample, have not been reached yet experimentally.
Recent very promising works, however, have shown new insights of this limit~\cite{balistico,Baringhaus}.
Beyond the sample-size dependency~\cite{Xu}, different arrangements of 
atomic vacancies, carbon isotopes, and distinct boundary conditions have also a strong impact  
on the phonon transport~\cite{Chen,Hu,Jiang,Zhang4,Haskings,Pop2}. 
Some results of these effects were already known from studies on 
carbon nanotubes~\cite{bookCNT, Mingo1,MingoCNT} and have been extended to graphene nanoribbons (GNRs)~\cite{Wu, pablo1}. 
In general, all these studies were focused mainly on the reduction of the total thermal conductivity 
leaving the microscopical details somehow unattended.
 
In this work, we demonstrate 
that it is possible to use adequate configurations of boundary conditions, 
antidot arrangements and isotopic barriers 
to tune the polarized ballistic phonon transmission in GNRs. 
We show that particular local atomic displacements can be controlled with the purpose of 
filtering incoming in-plane or out-of-plane phonon modes at specific energies.
In adittion to the tuning of the phonon transmission function T$(\omega)$,  
our findings have also important consequences to related problems of actual high interest that 
are the search for enhanced thermoelectric behaviors~\cite{Vaca}
and the improvement of mechanical properties by 
controlled defect creation~\cite{Mec}. 
%
 

{\it Model and Methods}. 
\label{sec1}
Interactions between carbon atoms 
inside GNRs were modelled with a harmonic  
fifth-nearest neighbor force-constant model (5NNFCM). 
Parameters of the 5NNFCM account for the radial bond-stretching, in-plane 
and out-of-plane tangential bond-bending interactions~\cite{Mohr}. 
This model has proven to describe phonon dispersions and elastic constants
of single and multi-layer graphene with excellent accuracy~\cite{Michel}.
It was also used recently by us to investigate the role of single atomic vacancies and boundary conditions in the
thermal transport properties of GNRs~\cite{pablo2}.
Here, we expand the study with special emphasis on the 
management of the polarized transmission functions.  

\begin{figure*}[th]
\includegraphics[trim = 0cm 0cm 0.1cm 0cm, clip,width=0.272\textwidth]{Comp_10barreras_seba-OK.eps}
\hspace{0.03cm}
\includegraphics[trim = 0cm 0cm 0cm 0cm, clip, width=0.266\textwidth]{Grafica_uno_corr-OK.eps}
\hspace{0.03cm}
\includegraphics[trim = 0cm 0cm 0.5cm 0cm, clip, width=0.43\textwidth]{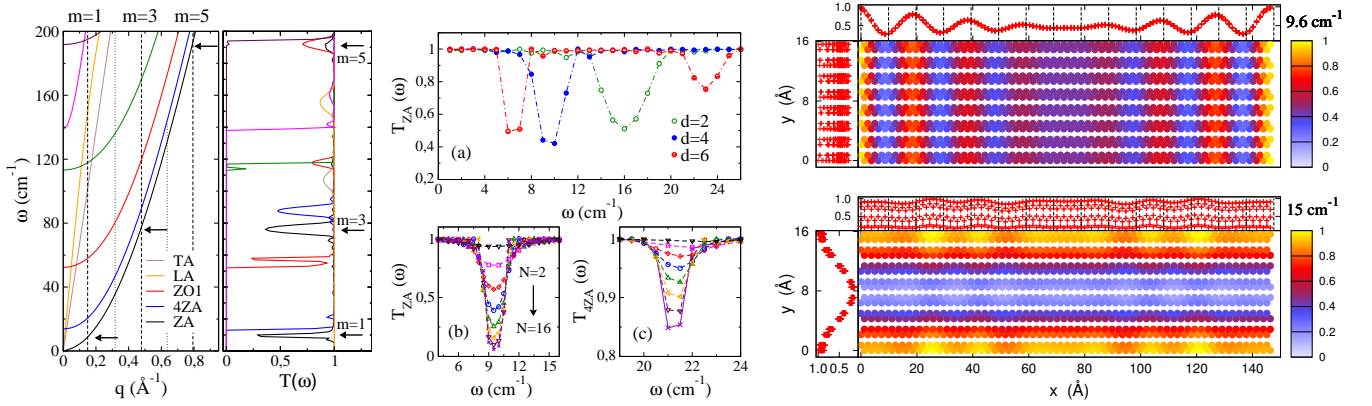}
\caption{
Isotopic barriers. 
Left: Phonon relation dispersions of a free-edge homogeneous ZGNR and polarized transmission fucntions T$(\omega)$ 
for $N=10$ isotopic barriers separated by $d=4$.
Center: (a) Variation of T$_{\mbox{\scriptsize{ZA}}}(\omega)$  against the barrier distance $d$=2,4,6 for $N$=8. 
Dependence with $N$=2,4,...,16 for the first dip of (b) T$_{\mbox{\scriptsize{ZA}}}(\omega)$ 
and (c) T$_{\mbox{\scriptsize{4ZA}}}(\omega)$ for $d=4$.
Right: LDOS profile at $\omega$=9.6 (1st dip) and 15 $\mbox{cm}^{-1}$ for $N=8$ and $d=4$.
}
\label{fig1}
\end{figure*}

Zigzag graphene nanoribbons (ZGNRs) were defined in the usual way
respecting the hexagonal lattice and the edge-shapes at the boundaries. Definitions of free- and supported-edge boundary
conditions and calculations of the contributions to T$(\omega)$ were performed using the formalism presented in Ref.~[\onlinecite{pablo2}].
The following systems are investigated in detail:
(i) C$^{13}$ isotopes were modelled by changing the atomic mass of carbon atoms
keeping unaltered the magnitude of the forces; 
(ii) atomic vacancies were introduced by switching off the interatomic interactions with the missing atoms 
and (iii) partially supported-edges were defined by fixing certain carbon atoms outside the ribbon.
Although not shown here, similar results were obtained for armchair GNRs. 

In the absence of inhomogeneities, the phonon spectrum 
can be obtained by numerical diagonalization of the dynamical matrix~\cite{Michel}.
The phonon spectrum of an homogeneous 
free-edge ZGNR contains four acoustical phonon modes,
i. e., one extra acoustical mode as compared to bulk graphene.
In the long wavelength regime the in-plane longitudinal (LA) and transversal (TA) modes 
exhibit linear energy dispersion, the out-of-plane (ZA) mode has quadratic dispersion and
the so called fourth acoustic (4ZA) mode posses a linear energy dispersion~\cite{pablo2}.
Similar to what occurs in CNTs, as shown in Fig.~\ref{fig1} (left-panel), this last mode has a small size-dependent energy gap 
when it is obtained using a force constant model~\cite{droth,Yacobson,Huang, Yamada, Guillen}. 
At higher energies, out-of-plane (ZO-nth) and in-plane optic modes are present. 
For supported-edge ZGNRs, all phonon modes develop an energy gap due to the 
breakdown of the translational symmetry~\cite{pablo2}.

The calculation of the energy-dependent transmission function T($\omega$) was performed 
by using the non-equilibrium Green's function formalism
within the conventional Landauer method~\cite{Zhang3, WangAgarwalla, bookZhang}.
Anharmonic terms and electron-phonon interactions were neglected. 
As described elsewhere, the central region containing antidots, isotopes or supported-edge zones,
was connected to two homogeneous semi-infinite contacts at different temperatures. 
Surface Green's functions (SGF) were calculated iteratively by using the decimation technique~\cite{Zhang1,Lopez}.
The polarized incident components of T$(\omega)$  were identified by proper rotation of the SGF to the 
basis of normal phonon modes~\cite{specific}.


{\it Results}.
\label{sec2}{\it A) Out-of-plane phonon modes}.
The quadratic phonon dispersion displayed by the out-of-plane ZA phonon mode  
at low energy entails important consequences in the temperature-dependent properties 
of the graphene lattice.
For instance, the ZA mode is responsible for the subtle lattice thermal 
contraction which takes place at intermediate temperatures~\cite{mounet}. 
%
The smaller group velocity $\nu_{n}(\omega)=\partial\omega/\partial q$ of the out-of-plane mode ($n$=ZA)
in the long wavelength regime as compared to the in-plane modes have also important effects on the phonon transport.
Given a fixed wave-vector, in-plane and out-of-plane incident waves 
will propagate at different speeds. Consequently, they interact 
differently with any obstacle (atomic vacancies, carbon isotopes, any kind of impurity, etc.) 
encountered on its way.
If the obstacle is localized (extent limited to a small region in space)
fast moving waves will transit almost without being affected, but slow moving waves
will suffer partial reflection. 

This effect can be more clearly seen in 
the analytical expression for the amplitude mode transmission $t^{RL}_{n}(\omega)$ 
obtained by using a mode-matching technique~\cite{wang} in which $t^{RL}_{n}(\omega)=
i2\omega \{(E^+_R)^{-1} G_{N+1,0} [(E^-_L)^T]^{-1}\}_{n} \nu^L_n(\omega)$, where
$\nu^L_n$ is the group velocity of the incident wave of mode $n$, $G_{N+1,0}$ 
is the corner element of the retarded Green's function for the junction part 
and $E^+_R$ and $E^-_L$ are the matrices formed by the column 
eigenvectors for the reflected normal modes of the left and right leads, respectively. 
In the following we will show that by using specific configurations of isotope 
and antidot structures this phenomenon can be exploited in order to 
filter slow moving out-of-plane phonons at low energy.

\begin{figure*}[th]
\includegraphics[width=0.32\textwidth]{Comp_10nanoporos_seba-OK.eps}
\hspace{0.2cm}
\includegraphics[width=0.24\textwidth]{Grafica_dos_corr-OK.eps}
\hspace{-1.48cm}
\includegraphics[width=0.43\textwidth]{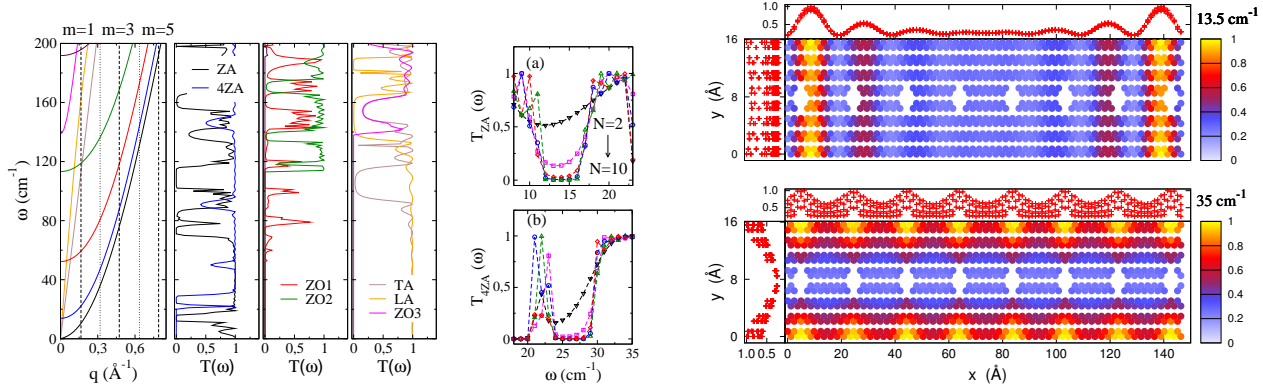}
\caption{
Antidot lattice. 
Left: Phonon relation dispersions of a free-edge homogeneous ZGNR and T$(\omega)$ for 
$N$=10 periodic antidots and $d$=4.
Center: Dependence with $N$=2,4,...,10 for the first dip of (a) T$_{\mbox{\scriptsize{ZA}}}(\omega)$ 
and (b) T$_{\mbox{\scriptsize{4ZA}}}(\omega)$ for $d$=4.
Right: LDOS profile at $\omega$=13.5 (1st dip) and 35.0 $\mbox{cm}^{-1}$ for $N$=8 and $d$=4.
}
\label{fig2}
\end{figure*}

We start by analyzing the effect of $N$ finite periodic isotopic barriers positioned along the ZGNR. 
Each barrier consists of four consecutive columns of C$^{13}$ isotopes, which are 
separated from each other by $d$ columns of C$^{12}$ atoms. 
The small mass variation ($\sim$ 8.34$\%$) between C$^{12}$ and C$^{13}$ has 
a negligible impact on the thermal transmission when there are only a few random distributed isotopes (diluted limit).
Only in case of large concentration of isotopes the thermal transmission becomes significantly distorted.
In Fig.~\ref{fig1} (left-panel) we show the low energy behavior of T$(\omega)$ 
for $N$=10 and $d$=4. 
As observed, each single phonon mode transmission presents a few visible dips.
The energies at which these dips are present for each phonon mode $n$ can be estimated as $\omega_{m}^n(q=m\pi/L)$  
where $m$ accounts for the $m$-th dip, $L=(4+d)\sqrt{3}a$
is the distance between isotopic barriers and $a$=1.42$\AA{}$ the bond length between carbon atoms.
For a better comprehension we added arrows in order to identify the $m$=1,~3 and 5 dips for the ZA mode on the plot. 
This situation resembles those for the transmission 
coefficients of weighted strings~\cite{gri1,stia1}. Accordingly,
the intensity of the dips is afffected by an extra modulation which depends on the barrier width~\cite{gri2}.
In this particular case, the modulation produce a large attenuation of the $m$=2 and 4 dips.

The shift of the first dip ($m$=1) of T$_{\mbox{\scriptsize{ZA}}}(\omega)$ against the inter-barrier distance $d$
is displayed in Fig.~\ref{fig1} (center, a). 
The estimated energies $\omega^{\mbox{\scriptsize{ZA}}}_1=16.7$, 9.6  and $6.6$ cm$^{-1}$ 
for $d$=2,4,6, respectively, are consistent with the values observed.
In Figs.~\ref{fig1} (b and c), we show the increasing intensity of the first dip
of T$_{\mbox{\scriptsize{ZA}}}(\omega)$ and T$_{\mbox{\scriptsize{4ZA}}}(\omega)$ 
for increasing $N$, with fixed $d$=4.
%
%
Notice on the contrary that in-plane mode transmissions T$_{\mbox{\scriptsize{TA}}}(\omega)$ 
and T$_{\mbox{\scriptsize{LA}}}(\omega)$,  remain unity (perfect transmission) up to 
large energies (109.1 and $158.6$ cm$^{-1}$, respectively).
 
A better microscopic understanding of $T(\omega)$ is obtained by analyzing the
local density of states (LDOS) of the atomic displacements, as shown in Fig.~\ref{fig1} (right panel).
At $\omega=9.6$ cm$^{-1}$, where the first dip of T$_{\mbox{\scriptsize{ZA}}}$ appears, carbon atoms are 
confined to move between the first two closest barriers 
and the propagation of the incident wave is strongly inhibited. 
Observe that in the center of the periodic arrangement, LDOS is $<$0.3.
%
%
At a larger energy, $\omega=15$ cm$^{-1}$,  the incident wave exhibits 
almost perfect transmission through the set of barriers. 
Now, the active contribution of the edge-localized 4ZA mode produces larger 
displacements at both edges of the ribbon where the LDOS is approximately equal to one.
%

We now proceed with the analysis of T$(\omega)$ for a finite array of $N$ periodic antidots.
This kind of systems are currently of high interest due to their 
expected large thermoelectric characteristics\cite{Jouho,pattern,physicaA,nano1}. 
Here, each antidot has been defined by eliminating a group of 6 carbon atoms 
and the distance between them is kept as in the case of isotopic barriers
in order to obtain comparable behaviors. 

In Fig.~\ref{fig2} (left-panel) we show therefore the 
low energy behavior of T$(\omega)$ for $d$=4 and $N$=10. 
Similarly as above, an overall analogous behavior with the gradual apparition of dips is found. 
Here however, the antidot structure produces stronger perturbations 
in the atomic displacements and hence its effect on T$(\omega)$ is considerably larger.
%
%
%
In Fig.~\ref{fig2} (center-panel) it can be observed that by increasing $N$ also here 
the intensity of dips becomes larger and already for $N$=8 one gets
T$_{\mbox{\scriptsize{ZA}}}(\omega)$ and T$_{\mbox{\scriptsize{4ZA}}}(\omega)$ 
$\approx 0$ at the first dip. 
%
The LDOS corresponding to the first dip of the ZA mode at $13.5$ cm$^{-1}$ shown in Fig.~\ref{fig2} (right) 
shows again localized vibrations restricted to the area enclosed by the first set of antidots.
Then, for $\omega=35~ \mbox{cm}^{-1}$ where T$_{\mbox{\scriptsize{ZA}}}\approx 1$, 
the vibrational pattern is similar to the case analyzed previously where the 4ZA mode becomes active
but there exist now an additional modulation of the displacements 
in the short direction of the ribbon due to the presence of the antidots.

\begin{figure*}[th]
\vspace{0.2cm}
\includegraphics[width=0.37\textwidth]{Comp_10barreras_mixtas_seba-OK.eps}
\hspace{0.10cm}
\includegraphics[width=0.12\textwidth]{Grafica_tres_corr-OK.eps}
\hspace{0.10cm}
\includegraphics[width=0.40\textwidth]{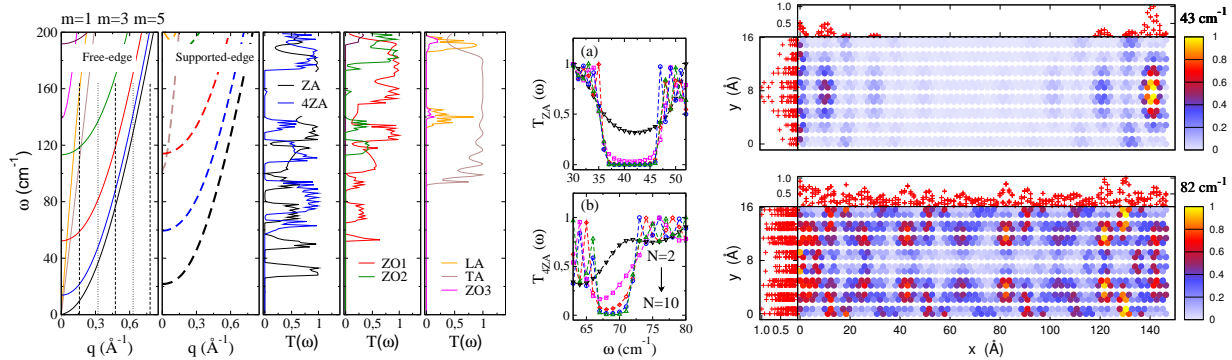}
\caption{
Partially supported-edges. 
Left: Phonon relation dispersions of a free- (continuous) and supported-edge (dashed lines) 
homogeneous ZGNR and T$(\omega)$ for $N$=10 stripes and $d$=4.
Center: Dependence with $N$=2,4,...,10 for the first dip of (a) T$_{\mbox{\scriptsize{ZA}}}(\omega)$ 
and (b) T$_{\mbox{\scriptsize{4ZA}}}(\omega)$ for $d$=4.
Right: LDOS profile at $\omega=$43.0 (1st dip) and 82.0~$\mbox{cm}^{-1}$ for $N$=8 and $d$=4.
}
\label{fig3}
\end{figure*}

{\textit{B) In-plane phonon modes}}.
In-plane LA and TA modes are characterized by their conventional acoustic 
linear dispersions at low energy. Their group velocities, estimated by using the 5NNFCM potential, are  
$\nu_{\mbox{{\scriptsize TA}}}=14.3\times10^5 \mbox{cm s}^{-1}$ and 
$\nu_{\mbox{{\scriptsize LA}}}=23.1\times10^5 \mbox{cm s}^{-1}$ for 2D graphene~\cite{Michel}.
These values are relatively large when compared to other materials\cite{review-phonons}. 
We show here that a more effective way of filtering these modes is, instead of using localized 
perturbations which only affect modes with low group velocity~\cite{wang}, by adopting partially supported-edges.
%
Within this setup, energy gaps at $q$=0 are opened for all the phonon modes in the spectra. 
There are, however, differences in the magnitude of the gaps being significantly larger 
for the in-plane phonon modes as compared to the out-of-plane ones~\cite{pablo2}.

In what follows, we explore the effect of partially supported-edge ZGNRs on T$(\omega)$. 
The configuration considered consists of $N$ periodic stripes, each of which consists of 
4 columns of atoms where only those atoms laying at the edges are kept fixed.
In this case a combined reduction effect for in-plane and out-of-plane transmissions is expected 
at low energy.

To facilitate the understanding of the results in 
Fig.~\ref{fig3} (left-panel), we show also the phonon dispersion relations 
for an homogeneously supported-edge (dashed lines) ribbon, 
which displays the energy gaps~\cite{pablo2} mentioned above,
together with T$(\omega)$ for the case $N$=10 and $d$=4. 
%
%
The main difference is that now the transmission functions of the acoustical modes T$_{\mbox{\scriptsize{ZA}}}$,
T$_{\mbox{\scriptsize{LA}}}$, T$_{\mbox{\scriptsize{TA}}}$ and T$_{\mbox{\scriptsize{4ZA}}}(\omega)$ are zero  
until the energy of the incident waves reaches the values observed in Fig.~\ref{fig3} (left-panel).  
These values are very close to the corresponding gaps of the first two out-of-plane modes and the first in-plane
transversal mode of the supported-edge ZGNR.
In this sense, note that while out-of-plane modes become active at relative small energies, 
T$_{\mbox{\scriptsize{TA}}}(\omega)$ and T$_{\mbox{\scriptsize{LA}}}(\omega)$ 
remain zero up to $\omega\sim92$ and $133$ cm$^{-1}$, respectively. 
The identification of dips produced by the periodic arrangement 
of stripes is now no longer simple as it was in the case of isotopic barriers.
However,  still now the first visible dips become more pronounced by increasing $N$ 
for T$_{\mbox{\scriptsize{ZA}}}(\omega)$ and T$_{\mbox{\scriptsize{4ZA}}}(\omega)$
as can be observed in Fig.~\ref{fig3} (center-panel).

Finally, in Fig.~\ref{fig3} (right-panel) we show the LDOS 
at $\omega$=43  and $82$ cm$^{-1}$ for $N$=8 and $d$=4.
In the first case, the atomic displacements become gradually suppressed towards the center of the ribbon.
Note also the nearby zero LDOS for edge-atoms at the supported parts of the ribbon.
Then, for $\omega=82$ cm$^{-1}$ the situation is clearly more complex and the atomic displacements show traces of localization.
Inside the supported-edge stripes, the LDOS displays the pattern of the second out-of-plane mode of supported-edge ZGNRs~\cite{pablo2},
while inside the free-edge stripes the LDOS shows a partial contribution of the 4ZA mode. 


{\it Conclusions}.\label{sec4}
We have investigated the tuning of polarized ballistic thermal transmission functions in graphene nanoribbons by 
using different boundary conditions and proper arrangements of antidot structures and C$^{13}$ isotopic barriers.  
We demonstrated that by adopting adequate configurations it is possible to tune the ballistic 
phonon transport by controlling the frequency and the magnitude of the transmission functions T$(\omega)$. 
When the width (W) of the GNR becomes larger, the energy gaps of the phonon modes at $\vec{q}$=0 
are expected to scale as $\sim$ 1/W and 1/W$^2$, for in-plane and out-of-plane modes, 
respectively~\cite{pablo2}. Therefore, the overall trends are expected to remain valid. 

In addition to the systems proposed here, similar effects 
could be achieved by other means such as using proper configurations 
of adatoms like hydrogen\cite{sandeep1}, fluorine\cite{sandeep2}, etc.
In the experimental setup, supported boundaries can be obtained in several ways.
The ribbon could be deposited by its edges over a substrate. 
In this case one expects van der Waals type interactions 
acting between distinct atoms and restricting the movement of edge-lying atoms.   
A good candidate with a very low lattice mismatch is h-BN\cite{seba2,novo1}.  
A stronger constriction can be obtained by adding a metallic material 
on top of the edges where covalent bonds are expected to form. 
Here, however, one may have extra contributions to the thermal transport 
from the free electrons in the metallic contacts. 




{\it Acknowledgments}. Discussions with C. E. Repetto, C. R. Stia and K. H. Michel are gratefully acknowledged.
This work was partially supported by the Flemish Science Foundation (FWO-Vl) 
and PIP 11220090100392 of CONICET (Argentina). 
We acknowledge funding from the FWO~(Belgium)-MINCyT~(Argentina) collaborative research
project. 
\label{agradecimientos}


\end{document}